\documentclass[%
 reprint,
superscriptaddress,
 amsmath,amssymb,
 aps,
 prl,
]{revtex4-2}
\usepackage{xcolor}
\usepackage{graphicx}
\usepackage{dcolumn}

\begin{document}
\preprint{APS/123-QED}

\title{Search for fractionally charged particles with CUORE}

\author{D.~Q.~Adams}
\affiliation{Department of Physics and Astronomy, University of South Carolina, Columbia, SC 29208, USA}

\author{C.~Alduino}
\affiliation{Department of Physics and Astronomy, University of South Carolina, Columbia, SC 29208, USA}

\author{K.~Alfonso}
\affiliation{Center for Neutrino Physics, Virginia Polytechnic Institute and State University, Blacksburg, Virginia 24061, USA}

\author{F.~T.~Avignone~III}
\affiliation{Department of Physics and Astronomy, University of South Carolina, Columbia, SC 29208, USA}

\author{O.~Azzolini}
\affiliation{INFN -- Laboratori Nazionali di Legnaro, Legnaro (Padova) I-35020, Italy}

\author{G.~Bari}
\affiliation{INFN -- Sezione di Bologna, Bologna I-40127, Italy}

\author{F.~Bellini}
\affiliation{Dipartimento di Fisica, Sapienza Universit\`{a} di Roma, Roma I-00185, Italy}
\affiliation{INFN -- Sezione di Roma, Roma I-00185, Italy}

\author{G.~Benato}
\affiliation{Gran Sasso Science Institute, L'Aquila I-67100, Italy}
\affiliation{INFN -- Laboratori Nazionali del Gran Sasso, Assergi (L'Aquila) I-67100, Italy}

\author{M.~Beretta}
\affiliation{Department of Physics, University of California, Berkeley, CA 94720, USA}

\author{M.~Biassoni}
\affiliation{INFN -- Sezione di Milano Bicocca, Milano I-20126, Italy}

% ...


\date{\today}

\begin{abstract}
The Cryogenic Underground Observatory for Rare Events (CUORE) is a detector array comprised by 988 5$\;$cm$\times$5$\;$cm$\times$5$\;$cm TeO$_2$ crystals held below 20 mK, primarily searching for neutrinoless double-beta decay in $^{130}$Te. Unprecedented in size amongst cryogenic calorimetric experiments, CUORE provides a promising setting for the study of exotic through-going particles. Using the first tonne-year of CUORE's exposure, we perform a search for hypothesized \textit{fractionally charged particles} (FCPs), which are well-motivated by various Standard Model extensions and would have suppressed interactions with matter. No excess of FCP candidate tracks is observed over background, setting leading limits on the underground FCP flux with charges between $e/24-e/5$ at 90\% confidence level.  Using the low background environment and segmented geometry of CUORE, we establish the sensitivity of tonne-scale sub-Kelvin detectors to diverse signatures of new physics.

\end{abstract}

\maketitle

Charge quantization remains an unsolved mystery of the Standard Model (SM). Since the measurement of the electron charge quantum $e$ by Millikan and Fletcher~\cite{millikan}, the charges of all known elementary particles have been empirically found to follow a simple rule: their charges are all either zero, $\pm e, \:\pm \frac{1}{3}e,$ or $\pm \frac{2}{3}e$. Of these, only particles with integer electron charges ($q = ne$ for $n = 0, \pm 1$) have been observed as free particles~\cite{pdg}. 

Yet, there is no \textit{a priori} reason that the charge must be quantized. Existing explanations, such as Dirac quantization through magnetic monopoles~\cite{dirac} or Grand Unified Theories (GUTs)~\cite{pdg}, are yet to be confirmed, and the ever-expanding theoretical landscape has introduced many promising extensions to the Standard Model that permit free fractionally-charged particles (FCPs). These candidates can arise from theories with non-standard charge quantization \cite{chargequant1, chargequant2}, hidden sector couplings to vector bosons~\cite{vectorboson_DS}, or additional gauge groups~\cite{supersym,gauge_mixing, gauge2} and may appear in the form of unbound quarks~\cite{free_quarks, free_quark2} or novel leptons~\cite{lepton, lepton2}.

Fractionally charged particles (also known as \textit{lightly ionizing} particles or \textit{millicharged} particles when $q\ll e$) are parameterized by charge $q = e/f$ with $f>1$, and would present distinct experimental signatures due to their suppressed ionization energy losses relative to known charged particles.  FCPs have been the subject of an extensive experimental effort over the past several decades~\cite{perl_searches_2009}, spanning particle accelerators~\cite{LHC, cmscollaboration2024search_suep, PhysRevLett.124.131801}, balloon and space satellite experiments~\cite{DAMPE}, and bulk matter searches~\cite{alvis_first_2018, alkhatib_constraints_2021, earthbound}. Direct detection searches are frequently non-specific and model-independent and assume that fractionally charged particles may be present within cosmic radiation, which can be mapped to model-specific production mechanisms~\cite{Recast,PhysRevD.103.075029,wu2024searching}.  In this Letter, we report on a search for an underground flux of fractionally charged particles in the range $f=2-24$ with the CUORE experiment.

CUORE, the Cryogenic Underground Observatory for Rare Events, is a tonne-scale millikelvin experiment located underground at the Laboratori Nazionali del Gran Sasso (LNGS) in Italy, with the primary purpose of searching for neutrinoless double beta decay ($0\nu\beta\beta$) in $^{130}\text{Te}$~\cite{cuorecollaboration2024nu}. The experiment consists of an array of 5$\;$cm$\times$5$\;$cm$\times$5$\;$cm TeO$_2$ crystals at $\sim$10 mK housed within the CUORE cryostat~\cite{CUORE-Cryostat}. Each crystal detector is coupled to a neutron transmutation doped germanium thermistor to read out thermal pulses produced by energy depositions in the crystal~\cite{Haller1984}. 

CUORE’s highly segmented geometry, comprised of 988 crystal detectors arranged in 19 four-column vertical towers, aids background-rejection for $0\nu\beta\beta$ searches through anti-coincidence between channels~\cite{cuorecollaboration2024nu}, and has been used to search for rare nuclear decays of $^{120}$Te and $^{130}$Te with signatures distributed over 2-3 crystals~\cite{PhysRevC.105.065504,excited_states_CUORE}. We extend CUORE's capability to reconstruct detector-wide signatures of new physics, and establish that cryogenic bolometric detectors have reached sufficient scale to reconstruct through-going particle tracks.

FCPs would interact with the CUORE detector primarily through ionization energy loss, leaving linear tracks across multiple crystals in the detector. The stopping power is fainter than a $q=e$ minimally ionizing particle, leaving a distinct signature of new physics. Similar to the treatment in~\cite{alvis_first_2018}, the flux $\Phi(f)$ of FCPs through the detector may be expressed as 
\begin{equation}\label{eq:flux}
\Phi(f) = \dfrac{n^{\text{Sig}}}{T_{\text{livetime}}\cdot(A \Omega)_{\text{selection}}\cdot \epsilon_{\text{cluster}}},
\end{equation}
where $n^{\text{Sig}}$ is the total number of signal candidates (observed or undetected limits), $(A \Omega)_{\text{selection}}$ is CUORE's acceptance to FCPs coming from the inherent detector physics/geometry and analysis selections,  $\epsilon_{\text{cluster}}$ is a global efficiency term from the probability of successfully temporally clustering together all detector events induced by an FCP, and $T_{\text{livetime}}$ is the total detector livetime of the search. Values for $n^{\text{Sig}}$ and $(A \Omega)_{\text{selection}}$ will generally be $f$-dependent.

\begin{figure}[h!]
    \centering
    \includegraphics[width=.4\textwidth]{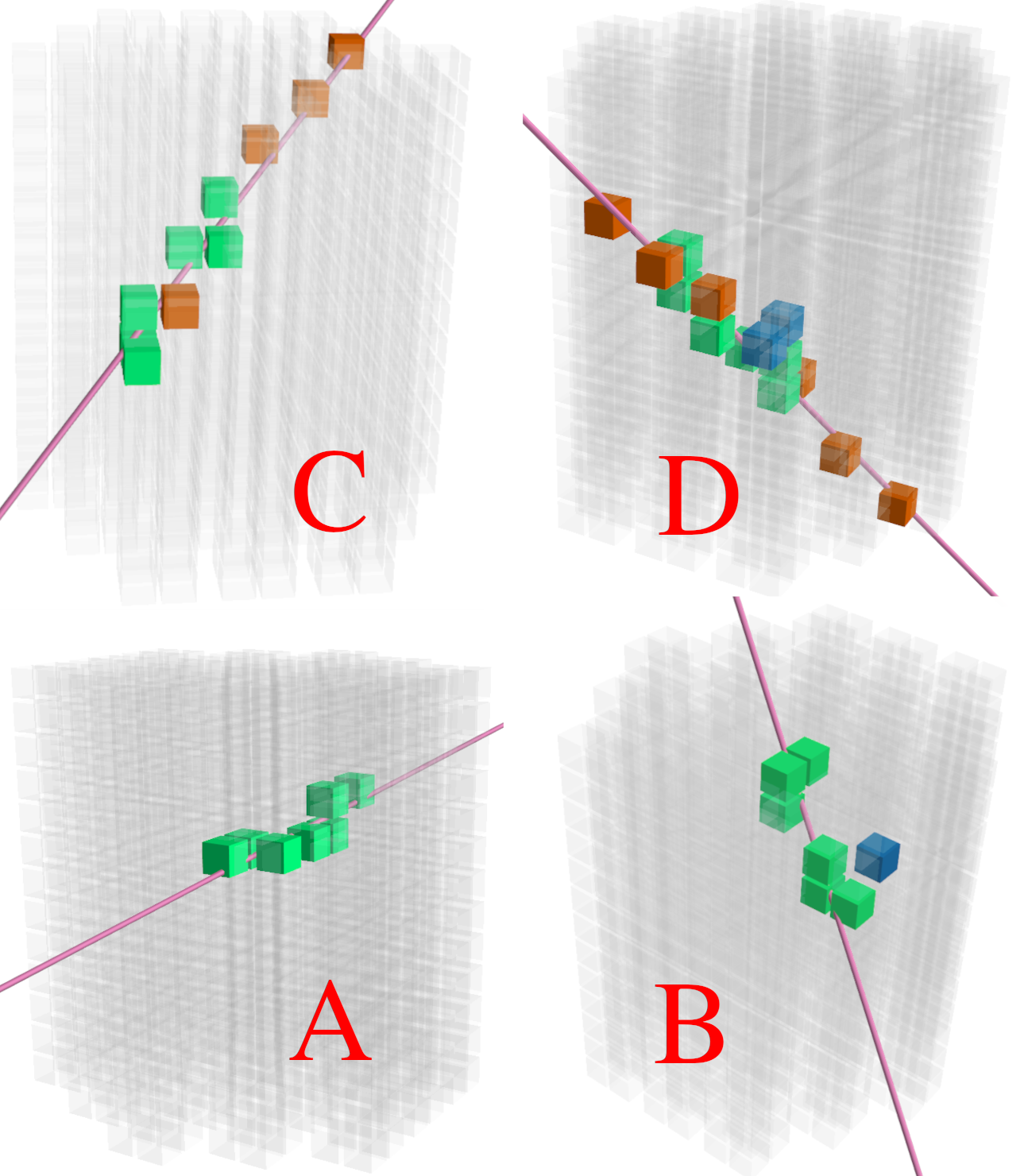}
    \caption{ Four example track-fit clusters observed in data as they appear within the CUORE detector array. The best-fit tracks are shown in purple, while channels which are both intersected by the best-fit track and register an event are shown in green. ``Missing'' channels corresponding to channels which register an event but are not intersected by the best-fit track are shown in blue, while ``extra'' channels which are intersected by the best-fit track without a corresponding observed event are shown in red. Channels which are non-participating in the cluster or track-fit thereof are shown translucently. The four example tracks (A, B, C, D) are each representative of the ABCD categorizations based on the presence or absence of extra- and missing-channels exhibited by the fitted cluster topology, as used in the present analysis.}
    \label{fig:track}
\end{figure}

\textit{Methods and data selection.---}Physics data collected in CUORE are grouped into datasets with 1-2 month duration, bookended by the deployment of calibration sources about the detector. For this search, we use the first tonne-year of CUORE's exposure as described in~\cite{cuore_nature} collected over 15 datasets. We remove periods of time characterized by the anomalously high occurrence of correlated low-energy pulses in the detector which appear to be non-particle in origin. For searches in CUORE for signal signatures occurring in one or few crystals (such as $0\nu\beta\beta$), such periods can be excluded on a channel-wise basis to avoid losing exposure from channels which are operating nominally. This search targets broad detector-wide signatures across many crystals, thus we conservatively remove such periods altogether. This search considers a total exposure of 442.3 days of detector livetime.

We refer to reconstructed energy depositions within any given crystal as \textit{events}, and multiple events occurring concurrently within the detector as a \textit{cluster}. Events are triggered offline from continuously collected detector timestreams, which are then filtered for amplitude and energy estimation with methods described in~\cite{cuore_nature}. For this analysis, we consider events with reconstructed energies between 20 keV--6 MeV per crystal. We apply pulse-shape cuts using principal component analysis to reduce non-physical events and those with poor pulse reconstruction. We build clusters by grouping temporally related events with a boxcar filter, and identify clusters containing 6 or more contemporaneous events. We denote the number of events in a given cluster as the cluster \textit{multiplicity}, $\mathcal{M}$, corresponding to the number of crystals triggered in coincidence with each other. The filter window is tuned using estimates of CUORE's timing resolution derived from the inter-arrival time distribution of events within the detector. Selecting 80 ms, we find a detector-wide clustering efficiency of $\epsilon_{\text{cluster}}=94(1)\%$ for $\mathcal{M}\geq6$ events. 

Cosmic ray muons present a potential background to any linear track search in CUORE, which are expected to register in the detector with per-crystal energy depositions of $\mathcal{O}$(10-50 MeV) and may saturate the dynamic range of our read-out electronics~\cite{Arnaboldi_2018}. We veto against muon events by discarding clusters in coincidence with any saturated event or event greater than 10 MeV in energy. From Monte Carlo simulations of expected muon backgrounds, we find this self-veto criterion to be more than 99.9\% efficient at rejecting muons which directly pass through two or more crystals within the detector array. We additionally find that occasional thermal, vibrational or microphonic noise can produce manifestly colinear correlated events on a single detector column. We therefore reject clusters with at least 60\% of channels occurring within the same column.

Clusters passing these selections are then fit with a version of the multi-objective optimization (MOO) algorithm presented in \cite{yocum_muon_2022} tailored to FCP reconstruction. Fitting a track to a cluster of events within CUORE gives rise to two track consistency parameters: the number of \textit{extra} channels that are intersected by the fitted track but do not register an event, and the number of \textit{missing} channels that register energy but are not intersected by the fitted track. Clusters with low numbers of extra channels and missing channels (referred as \texttt{ExtraCh} and \texttt{MissingCh} respectively) are more indicative of a through-going particle track. Examples clusters and corresponding fitted tracks are shown in Fig.~\ref{fig:track}.

We modify the MOO algorithm to provide a maximum-likelihood estimate for the $f$-value that best reconstructs the observed energy depositions and track lengths. This enables us to directly search for an excess in the distribution of reconstructed-$f$ arising from track-fitting selected clusters. From each fitted cluster, we extract the tuple of fitted observables: ($f$, \texttt{ExtraCh}, \texttt{MissingCh}).

FCPs are simulated by incorporating the package developed by S. Banik and others \cite{banik_simulation_2020} into CUORE's \texttt{geant4} \cite{geant1,geant2,geant3} Monte Carlo (MC) detector model. The simulation treats FCPs as massive fermions with relevant electromagnetic loss processes correspondingly suppressed. We nominally consider a minimally-ionizing relativistic particle ($\beta \gamma=3$) of mass 100 GeV/$c^2$, but find that we are insensitive to differences in particle mass over the range of 100 MeV/$c^2$-1 TeV/$c^2$ and relativistic parameters between $\beta \gamma = 3-300$. We simulate FCPs at eight logarithmically-spaced charge values between $f=2-24$, sampled uniformly across a 15 meter radius disk centered above the detector.

Per dataset, we tune the output of these simulations to match detection inefficiencies as exhibited within the collected data, which can impact cluster reconstruction. We mimic channel-wise deadtime within the detector, reflecting when particular channels are not taking good physics data. We determine trigger probabilities with the synthetic data method presented in \cite{Gianvecchio’_2021}, which may be less than unity for energies below $\sim 40$ keV. Additionally, we consider a base-cut efficiency for whether an event will be well-reconstructed, as in \cite{cuore_nature}. Finally, we consider the event selection efficiency for pulse-cleaning cuts. Above 100 keV, we use the technique described in~\cite{cupid_mo} by counting the fraction of events within known $\gamma$ and $\alpha$ peaks passing and failing cuts. Below 100 keV, we determine our event selection efficiency using low-energy events arising from high-multiplicity electromagnetic cascades found in coincidence with through-going muon candidates.

\begin{figure}[h!]
    \centering
    \includegraphics[width=.45\textwidth]{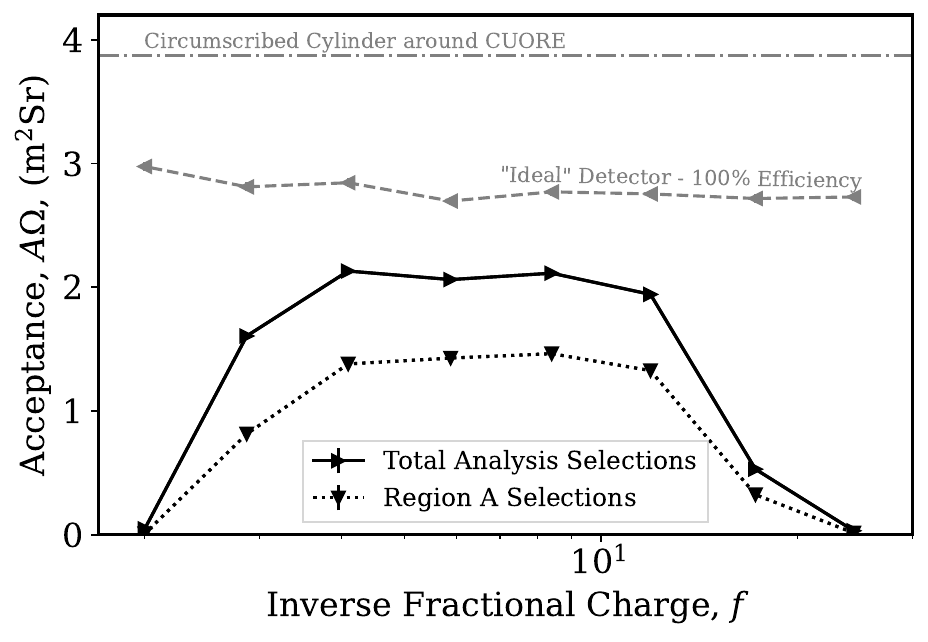}
    \caption{Acceptance of CUORE to isotropically downward-going FCPs, from Monte Carlo simulations of FCP interactions with the detector. Shown is the total acceptance entering into the ABCD template selections (solid black), along with the portion coming from clusters within Region A (dashed black). The falloff at high/low $f$-values reflects typical FCP energy depositions falling outside the 20 keV-6 MeV analysis selections of this search. Values are also shown for a reference cylinder circumscribed around the CUORE detector (dot-dashed grey), along with CUORE's acceptance to $\mathcal{M}\geq6$ clusters under idealistic conditions, i.e. assuming 100\% detection efficiency and without analysis selections on event energy (dashed grey).}
    \label{fig:exposure}
\end{figure}

From the efficiency-tuned MC output, we determine the area$\times$solid-angle geometric acceptance, $(A \Omega)$, of CUORE to isotropically downward-going FCPs, displayed in Fig.~\ref{fig:exposure}. Alternatively, if FCPs are assumed to follow a muon-like angular distribution within LNGS~\cite{AGLIETTA1994103}, these acceptance values are reduced by up to 50\%, reflecting the lower geometric and analysis acceptance of CUORE to a steeper muon-like flux.

\begin{figure}[h!]
    \centering
    \includegraphics[width=.45\textwidth]{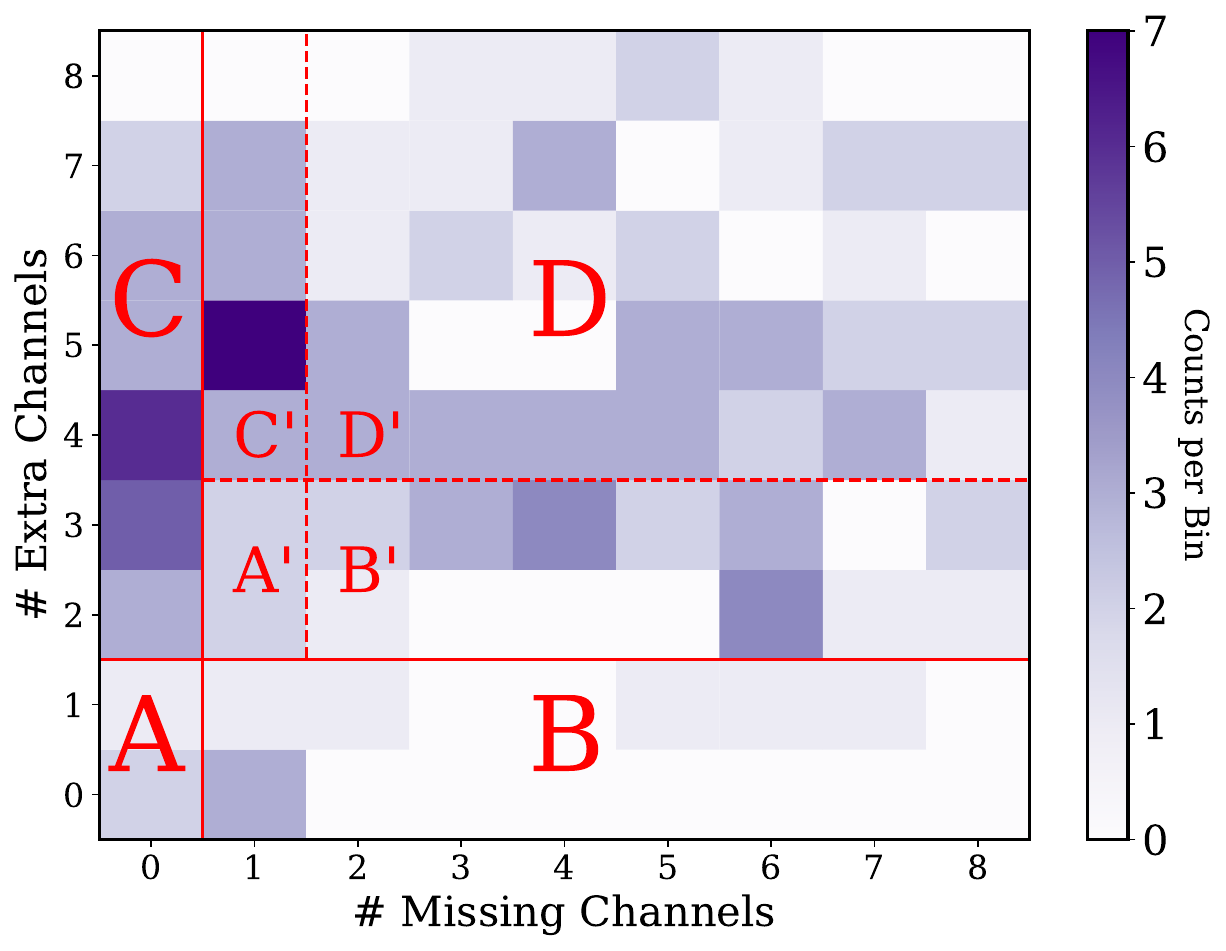}
    \caption{Observed clusters passing analysis selections binned within the plane of the number missing channels (\texttt{MissingCh}) and extra channels (\texttt{ExtraCh}) arising during track fitting. The selected ``ABCD'' regions are marked by solid red lines, while the validation A'B'C'D' subdivisions of Region D are marked by dashed red lines.}
    \label{fig:ABCD}
\end{figure}

\textit{Results.---}This search takes a data-driven approach to background estimation, using the ``ABCD'' method~\cite{ABCD_method} commonly employed for analyses at collider experiments (see for example~\cite{PhysRevD.44.29,atlas_abcd,cmscollaboration2024search_suep}). This technique leverages that the distributions of \texttt{ExtraCh} and \texttt{MissingCh} are found to be statistically independent for background events arising from non-tracklike sources, while being highly correlated and suppressed in number for clusters induced by FCPs. We bin fitted clusters by $f$ value into 3 bins logarithmically-spaced between $f=1-40$, and additionally define ``ABCD'' regions by the binary cuts in the number of extra- and missing-channels as shown by the solid divisions and labelling in Fig.~\ref{fig:ABCD}.

We utilize the following likelihood model to fit to both data and toy experiments:
\begin{equation}
\label{eq:lik}
    -2 \log \mathcal{L} = \sum_{i \in f\;\text{bins}}\; \sum_{j \in \{A,B,C,D\}} -2 \log \text{Pois} \left( k_{ij}; N_{ij} \right)
\end{equation}
where
\begin{equation}
\begin{split}
    N_{iA}&=\epsilon_{iA} n^{\text{Sig}} + n^{\text{Bgd}}_i\;,\\
    N_{iB}&=\epsilon_{iB} n^{\text{Sig}} + \tau_{iB} n^{\text{Bgd}}_i\;,\\
    N_{iC}&=\epsilon_{iC} n^{\text{Sig}}+ \tau_{iC} n^{\text{Bgd}}_i\;,\\
    N_{iD}&=\epsilon_{iD} n^{\text{Sig}} + \tau_{iB}\tau_{iC} n^{\text{Bgd}}_i\;.\\
\end{split}
\end{equation}
Here, $k_{ij}$ are the observed data in $f$-bin $i$ and region $j$, $n^{\text{Sig}}$ is our parameter of interest for the total number of signal events, $\epsilon_{ij}$ are normalized signal templates from our efficiency-tuned FCP MC simulations, while $n^{\text{Bgd}}_i$, $\tau_{iB}$, and $\tau_{iC}$ are $f$-bin specific background nuisance parameters encoding the ABCD behaviour into the model. Fitting is performed with the \texttt{iminuit} implementation \cite{iminuit} of the Minuit algorithm \cite{James:1975dr}. 

To reflect the degradation of signal candidate clusters due to accidental coincidences with uncorrelated background events within the detector, we move a corresponding 4.5\% of template weight from Region A to Region B and from Region C to Region D. Given our clustering method and selections, this reflects the probability of detector-wide pileup with an uncorrelated event, which can induce one or more additional missing channels when the resulting cluster is track-fit.

We perform our analysis with gradual unblinding of clusters within data. We start with the unblinding of clusters in Region D for model validation. Then we make available clusters within Region B and C for pre-unblinding sensitivity studies, and finally we unblind Region A with the analysis finalized.

We validate the ABCD model by examining Region D, which we expect to be comprised nearly entirely by background clusters. We further subdivide Region D into four validation inset regions as shown in Fig.~\ref{fig:ABCD}. We fit the validation regions to a background-only ABCD model, and compare the result with fits to toy-experiments drawn from the best-fit background parameters for the validation region. We find that the fitted value of $-2 \log \mathcal{L}$ lies in the 70$^{\text{th}}$ quantile of the sampling distribution from toy experiments, indicating good agreement between the ABCD model and the observed validation region. Additionally, a Pearson-$r$ test \cite{wasserman2010statistics} to data within Region D does not show evidence of correlation between the \texttt{ExtraCh} and \texttt{MissingCh} observables with an $r$-coefficient of -0.04 at $p=0.58$, further supporting the use of the ABCD model to describe background clusters.

We use a frequentist profile-likelihood ratio statistic to test between background-only and background-plus-FCP hypotheses ~\cite{asympt_tests}. We repeat this test at each simulated $f$-value separately and derive global test significances as corrected by a numerically-determined trial factor of 3.8 to account for the Look-Elsewhere Effect~\cite{LEE}. Since our data is in a low-statistics regime, we cannot rely on asymptotic approximations, and instead build sampling distributions from toy experiments to determine test-statistic thresholds and confidence brackets~\cite{Direct_DM}.

In constructing toy experiments, we vary background nuisance parameters according to the hybrid \textit{a posteriori} Highland-Cousins technique \cite{2207.14353}. We determine a multivariate normal (MVN) distribution for our background nuisance parameters from fits to the observed data salted with additional signal counts. For each toy experiment, we sample the MVN distribution to determine nuisance parameter values, and then Poisson-sample the resulting background template to obtain a toy background spectrum. Toy experiments with non-zero injected signal counts sample FCP signal templates accounting for inherent Poisson fluctuations, and the additional variance from finite Monte Carlo statistics.

\begin{figure}[h!]
    \centering
    \includegraphics[width=.45\textwidth]{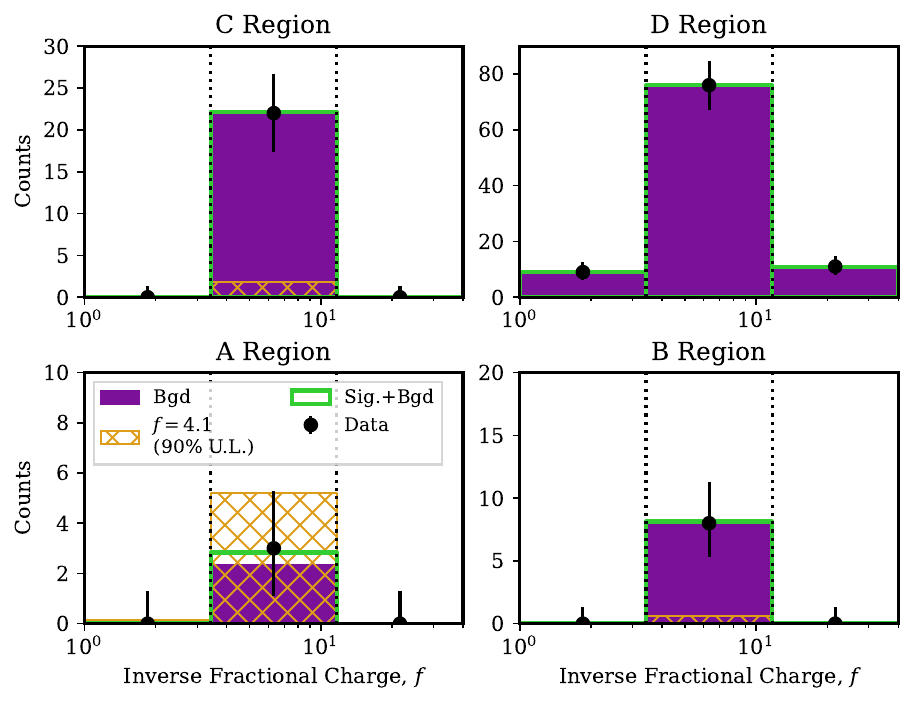}
    \caption{Observed cluster counts, binned by reconstructed $f$-value and ``ABCD'' subdivisions of \texttt{ExtraCh} and \texttt{MissingCh}. We also show the background-plus-FCPs best-fit to the data for $f=4.1$, compatible within $1\sigma$ with the absence of a signal. Displayed are the best-fit background component (solid purple), total signal-plus-background fit (green line), along with the signal template corresponding to the excluded 90\% upper-limit of 8.3 signal counts (hashed orange). }
    \label{fig:fit}
\end{figure}

\begin{figure}[h!]
    \centering
    \includegraphics[width=.45\textwidth]{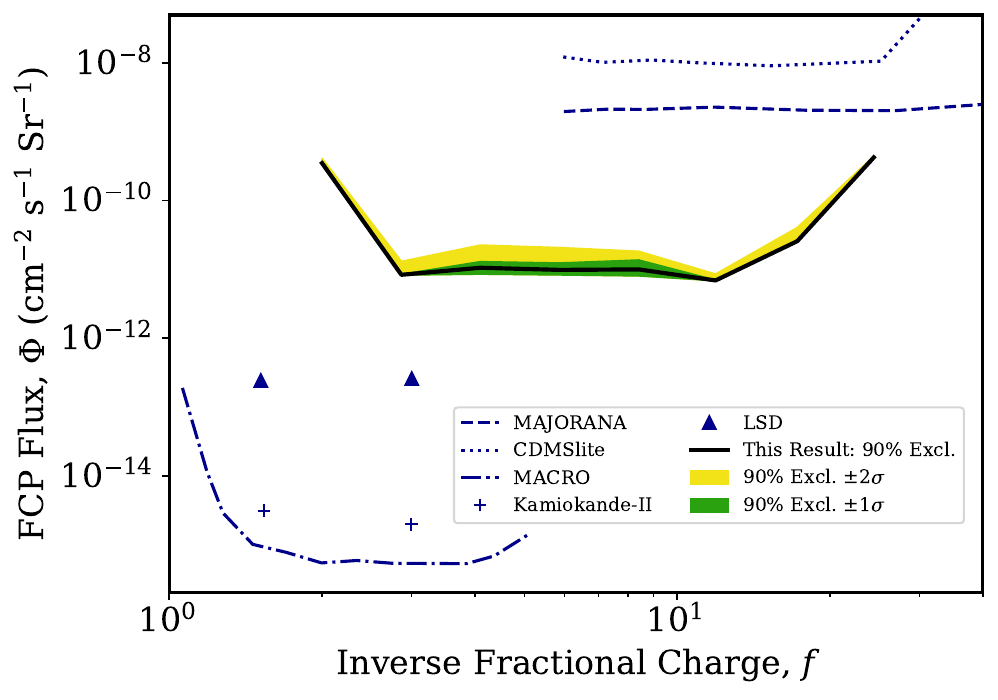}
    \caption{Observed exclusions at 90\% C.L. on the underground flux of FCPs as a function of inverse fractional charge, assuming a half-isotropic downward-going angular distribution. Also shown are the $\pm1\sigma$ and $\pm2\sigma$ ranges of expected exclusions under the background-only hypothesis, as derived from toy experiments. We compare with underground limits from other experiments \cite{alvis_first_2018,alkhatib_constraints_2021,macro_collaboration_search_2000,AGLIETTA199429,PhysRevD.43.2843}. }
    \label{fig:limits}
\end{figure}

Fitting to the observed data, we find no evidence for an excess of FCP-induced tracks, and find that the data is well-described by background-only fits across all tested values of $f$. Likelihood ratio test statistic values do not exceed 1$\sigma$ local (or global) significance in favor of FCPs across the values tested. The observed data, along with our best-fit ABCD reconstruction and signal exclusion at $f=4$, are shown in Fig.~\ref{fig:fit}.  We proceed to set upper-limits at 90\% confidence level on the observed number of signal counts using brackets built from fits to toy-experiments using a two-sided Neyman construction with Feldman-Cousins ordering~\cite{PhysRevD.57.3873,Direct_DM}.  We prevent ourselves from making exclusions stronger than would be made under an observation of zero signal counts with the method of Lokhov and Tkachov \cite{lokhov_tkachov}. We convert these $f$-dependent exclusions, which range between 5.1---8.3 signal counts, into limits on an underground flux of FCPs using Eq.~\ref{eq:flux}, which we display in Fig.~\ref{fig:limits} for a half-isotropic downward angular distribution. We find that these limits are world-leading amongst underground experiments for the range of inverse fractional charges between $f=5-24$ with a minimum exclusion of $\Phi < 6.9 \cdot 10^{-12}$ cm$^{-2}$ s$^{-1}$ Sr$^{-1}$ (90\% C.L.) at $f=11.9$, bridging a gap between historical general-purpose large-volume underground detectors \cite{macro_collaboration_search_2000,AGLIETTA199429,PhysRevD.43.2843}, and more contemporary searches targeting smaller charge values with reduced detector exposures \cite{alvis_first_2018,alkhatib_constraints_2021}. 

\textit{Conclusion.---}This search establishes new limits on the underground FCP flux, carving out significant new space across inverse charge parameters for relativistic particle species possibly present within cosmic radiation underground. More broadly, we demonstrate CUORE's capability to search for exotic detector-wide signatures of new physics. Analysis and processing techniques are under development to extend analysis thresholds and efficiencies in CUORE to lower energies, which would provide sensitivity to fractional charges more feebly-interacting than examined in this Letter. Similar searches in the forthcoming CUPID experiment~\cite{thecupidinterestgroup2019cupid} will benefit from finer detector segmentation, and additional detector information provided by the dual readout of heat and light signatures.

\begin{acknowledgments}
The CUORE Collaboration thanks the directors and staff of the Laboratori Nazionali del Gran Sasso and the technical staff of our laboratories. This work was supported by the Istituto Nazionale di Fisica Nucleare (INFN); the National Science Foundation under Grant Nos. NSF-PHY-0605119, NSF-PHY-0500337, NSF-PHY-0855314, NSF-PHY-0902171, NSF-PHY-0969852, NSF-PHY-1307204, NSF-PHY-1314881, NSF-PHY-1401832, and NSF-PHY-1913374; Yale University, Johns Hopkins University, and University of Pittsburgh. This material is also based upon work supported by the US Department of Energy (DOE) Office of Science under Contract Nos. DE-AC02-05CH11231 and DE-AC52-07NA27344; by the DOE Office of Science, Office of Nuclear Physics under Contract Nos. DE-FG02-08ER41551, DE-FG03-00ER41138, DE- SC0012654, DE-SC0020423, DE-SC0019316. This research used resources of the National Energy Research Scientific Computing Center (NERSC). This work makes use of both the DIANA data analysis and APOLLO data acquisition software packages, which were developed by the CUORICINO, CUORE, LUCIFER, and CUPID-0 Collaborations. The authors acknowledge Advanced Research Computing at Virginia Tech for providing computational resources and technical support that have contributed to the results reported within this paper. URL: https://arc.vt.edu/.
\end{acknowledgments}



\bibliography{bib}

\end{document}